\documentclass[twocolumn,pra,aps,superscriptaddress,showpacs]{revtex4}
\usepackage{graphicx}
\begin{document}
\title {Fermionic superfluidity and spontaneous superflows in
optical lattices}
\author{Shi-Jie Yang}
\affiliation{Department of Physics, Beijing Normal University,
Beijing 100875, China}
\author{Shiping Feng}
\affiliation{Department of Physics, Beijing Normal University,
Beijing 100875, China}

\begin{abstract}
We study superfluidity of strongly repulsive fermionic atoms in
optical lattices. The atoms are paired up through a correlated
tunneling mechanism, which induces superfluidity when repulsive
nearest-neighbor interactions are included in the Hubbard model.
This paired superfluid is a metastable state which persists for a
long time as the pair-broken process is severely suppressed. The
mean-field phase diagram and low energy excitations are investigated
in a square lattice system. Intriguingly, spontaneous superflows may
appear in the ground state of a triangular optical lattice system
due to antiferromagnetic frustration.
\end{abstract}
 \maketitle

\section{introduction}
Recently, the strongly correlated properties of ultracold atoms in
optical lattices have attracted great interests in
physicists\cite{Jaksch,Chin,Bloch}. The high tunability of the
interaction strength between atoms, as well as the easy manipulation
of the optical lattices makes the system realistically viable to
implement a quantum simulator\cite{Feynman}. The quantum atomic
gases in optical lattices make it possible to build a model system
so that we can explore the correlated properties in many-body
physics such as superconductivity, quantum magnetism, quantum
criticality, etc, and examine the related theoretical
models\cite{Jaksch2,Osterloh,Vidal}.

A recent experiment \cite{Winkler} showed that a couple of strongly
repulsive atoms occupying the same site of an optical lattice can be
stabilized by damping the single particle tunneling. The lifetime of
the pair increases significantly with the on-site repulsion $U$ of
the Hubbard model, which is quite intriguing as intuitively an
attractive force between particles is required to obtain a bound
state. In the presence of a periodic spatial potential, the energy
of a particle does not vary continuously but is restricted to
particular ranges of values. A pair of strongly repelling particles
can be stable because if it fell apart, the two isolated atoms would
ensure kinetic energies that fall in a forbidden band\cite{Fallani}.
Another experiment directly observed that for a pair of strongly
repulsive atoms in the optical lattice, single-particle tunneling is
severely suppressed by the requirement of energy conservation while
atom-pair co-tunneling is permitted through the second-order quantum
process\cite{Folling}. Although these experiments were performed on
bosonic atoms, it is conceivable that they can be applied to
fermionic atoms since no quantum statistics is involved.

The formation of metastable atom pair with repulsive interactions
was first proposed by A.F. Andreev in his study of diffusion of
impurities in quantum crystal\cite{Andreev,Andreev2}. The quantum
liquid of repulsive particle pairs in optical lattices has been
discussed by several
authors\cite{Petrosyan,Eckholt,Schmid2,Yang,Bak}. In a recent paper,
Rosch et al studied metastable superfluidity of repulsive fermionic
atoms in optical lattices\cite{Rosch}. Other authors attempted to
explore the possibility of counterflow superfluidity and to control
spin exchange interaction of two-species ultracold atoms in a
commensurate optical lattice\cite{Kuklov,Duan,Altman,Schmidt}. In
the fermionic superconductivity and superfluidity, a central concept
is pairing. In Bardeen-Cooper-Schrieffer (BCS) theory, electrons
pair up by an attractive interaction mediated by phonons of the
underlying crystals. The attraction between ultracold fermionic
atoms is provided by Feshbach resonances. Another pairing mechanism
comes from correlated hopping, which occurs in fermionic
tight-binding models. It has been suggested as a possible
explanation for the high-$T_c$ superconductivity
\cite{Ott,Orso,Kohl}.

In this paper we explore the fermionic superfluid (SF) in a system
consisting of repulsive atom-pairs in optical lattices. We deal with
a system partially filled by couples of spin-up and spin-down
fermionic atoms. The lattice site is either occupied by two atoms or
empty in the one-band Hubbard model. In the limit of the strong
on-site repulsion $U\gg t$, the pair-breaking tunneling is
suppressed as is revealed by the experiment [\onlinecite{Winkler}].
On the other hand, the atom-pairs may transport across the lattice
through the second-order quantum transition\cite{Folling}. As a
result, the atoms always move in pairs. The effective Hamiltonian is
obtained through the quantum perturbation theory, which is mapped to
an anisotropic \textit{antiferromagnetic} (AF) model instead of the
usual ferromagnetic model in the Bose-Hubbard model. The system
exhibits superfluidity when the nearest-neighbor (NN) repulsive
interactions are included. This paired superfluid is a metastable
state because the pair-breaking process is severely suppressed. We
investigate the mean-field (MF) phase diagram and low energy
excitations for a square lattice system. It exhibits a gapless mode
in the SF state and a gapful mode in the solid state.

The AF feature of the effective Hamiltonian may lead to an
interesting phenomenon when the underlying optical lattice is
triangular. As well-known, the ground state has a long-range $120^0$
N\'{e}el order. The variation of the azimuthal angles between the NN
spins corresponds to the phase modulation of the superfluid state,
which leads to a spontaneous superflow in the ground state due to AF
frustration. As a result, the system exhibits a pattern of
convection consisting of vortex-antivortex pairs.

The paper is organized as follows: In Sec.II we obtain an effective
Hamiltonian for strong on-site interactions by the second order
quantum perturbation approximation. In Sec.III we demonstrate the
superfliudity of the paired fermionic atoms by mapping the system to
the pseudospin-1/2 antiferromagnetic model. The phase diagram and
low energy excitations are calculated in the MF approximation. In
Sec. IV we explore the possible phenomenon of spontaneous superflows
of the paired atoms in a triangular lattice. A summary is included
in Sec.V.

\section{effective Hamiltonian}
We focus on the partial pair-filling system with $\nu<1$. For
sufficient low temperature and strongly repulsive on-site
interaction, the atoms will be confined to the lowest band which is
described by the Hubbard model \cite{Jaksch2},
\begin{equation}
\hat H=-\sum_{\langle ij\rangle \sigma}t_\sigma (f_{i\sigma}^\dagger
f_{j\sigma}+f_{j\sigma}^\dagger
f_{i\sigma})+U\sum_{i}n_{i\uparrow}n_{i\downarrow}+V\sum_{\langle
ij\rangle}n_{i}n_{j},\label{Ham}
\end{equation}
where $f_{i\sigma}$ ($f_{i\sigma}^\dagger$) is the annihilation
(creation) spin-$\sigma$ fermionic operator,
$n_i=n_{i\uparrow}+n_{i\downarrow}$ is the number operator with
$n_{i\sigma}=f_{i\sigma}^\dagger f_{i\sigma}$, and $t_{\sigma}$ is
the tunneling matrix element. $U>0$ is the on-site interacting
energy and $V>0$ is the NN interaction. In this work, we confine our
discussions to the case of $U\gg t_{\sigma}, V$.

Since the pair-breaking processes are suppressed, the single
particle hoping is eliminated in the second-order quantum
perturbation theory. The on-site Hubbard term is considered as the
the unperturbed Hamiltonain $H_0$. The hopping term is treated as
the perturbation $H_1$, which should be calculated to the second
order of $t_\sigma/U$ to avoid pair-breaking. The NN interaction
term commutes with the Hubbard term and will be included in the
effective Hamiltonian later. Using a generalization of the
Schriffer-Wolf transformation\cite{Hewson},
\begin{equation}
\bar H=H_0+\frac{1}{2}[S,H_1]+\frac{1}{2}[S,[S,H_1]]+\cdots,
\end{equation}
where $[S,H_0]=-H_1$ and $S^\dagger=-S$. Suppose $|\alpha\rangle$
are the degenerate paired states of the unperturbed $H_0$ with
energy $E_0$ and $|\beta\rangle$ are the pair-breaking intermediate
states of $H_0$ with $H_0|\beta\rangle=E_1|\beta\rangle$. Then
$E_1=E_0-U$. It should be emphasized that in the initial states
$|\alpha\rangle$ all atoms are paired up in the lattice sites while
in the intermediate states $|\beta\rangle$ only one pair of atoms is
breaking. We have $\langle\alpha |H_1|\alpha'\rangle=\langle\beta
|H_1|\beta'\rangle=0$ and $\langle\alpha
|S|\beta\rangle=\langle\alpha |H_1|\beta\rangle/U$. Disregarding the
intermediate states with more breaking atom-pairs which will involve
higher order of $t_\sigma/U$, the the second order quantum
perturbation Hamiltonian is then
\begin{equation}
\langle\alpha|H^{(2)}|\alpha'\rangle=\frac{1}{U}\sum_\beta\langle\alpha|
H_1|\beta\rangle\langle\beta|H_1|\alpha'\rangle.
\end{equation}
An alternative method of the degenerate quantum perturbation theory
can be found in Refs.[\onlinecite{Emery,Kuklov}].

The effective Hamiltonian is then,
\begin{widetext}
\begin{equation}
\hat H_{eff}=\frac{4t_\uparrow t_\downarrow}{U}\sum_{\langle
ij\rangle} (f_{i\downarrow}^\dagger f_{i\uparrow}^\dagger
f_{j\uparrow}
f_{j\downarrow}+f_{i\uparrow}f_{i\downarrow}f_{j\downarrow}^\dagger
f_{j\uparrow}^\dagger)+(V-\frac{t_\uparrow^2+t_\downarrow^2}{U})
\sum_{\langle ij\rangle}n_in_j+\frac{z(t_\uparrow^2+t_\downarrow^2)}
{U}\sum_i n_i,\label{eff1}
\end{equation}
\end{widetext}
where $z$ is the number of the NN sites. In the second-order quantum
perturbation, the intermediate virtual state $|\beta\rangle$ that
breaks the atom-pair has a lower energy ($-U$) than the initial and
final states $|\alpha\rangle$, which induce an attractive NN
interaction. In order to prevent the atom-pairs from congregation, a
moderate repulsive NN interaction $V$ is introduced in the original
Hamiltonian (\ref{Ham}) to overcome the induced attractive
interaction. The first term describes the pair-hopping, implying
that the spin-up and spin-down atoms transport together across the
lattice. The composite object behaves like a hardcore bosonic
molecule because the fermion pairs always hop together to their
nearest neighboring site and for each site only one pair of atoms is
allowed. It should be noted that the pairing in our work is of
s-wave type, where only the lowest single-particle band is
considered in the Hubbard model.

To study the MF properties of the system, it is convenient to map
the effective Hamiltonian (\ref{eff1}) to the spin representation
\cite{Kuklov,Isacsson} by defining $\textbf{S}_i$ as
$S_{ix}=(f_{i\downarrow}^\dagger f_{i\uparrow}^\dagger+
f_{i\uparrow}f_{i\downarrow})/2$, $S_{iy}=(f_{i\downarrow}^\dagger
f_{i\uparrow}^\dagger-f_{i\uparrow}f_{i\downarrow})/2i$, and
$S_{iz}=[S_{ix},S_{iy}]/i=(n_i-1)/2$,
\begin{widetext}
\begin{equation}
\hat H_{eff}=\frac{8t_\uparrow t_\downarrow}{U}\sum_{\langle
ij\rangle}
(S_{ix}S_{jx}+S_{iy}S_{jy})+4(V-\frac{t_\uparrow^2+t_\downarrow^2}
{U})\sum_{\langle ij\rangle}S_{iz}S_{jz}+2zV\sum_{i}S_{iz}.
\label{eff}
\end{equation}
\end{widetext}
The NN interaction $V$ also acts as an external magnetic field
exerting on the pseudo-spins.

In contrast to the ferromagnetic model \cite{Sachdev,Hebert} in the
usual hardcore Bose-Hubbard model, the effective Hamiltonian
(\ref{eff}) represents an anisotropic AF model, where there are
several competitive phases dependent on the pair-filling $\nu$ as
well as $V$ and $t_\sigma$. Rosch et al obtained a ferromagnetic
model by making a particle-hole transformation for the down
spins\cite{Rosch}. At the MF level, we minimize the energy at fixed
$z$-polarization or pair-filling. Suppose the classical spins
$\textbf{S}_i$ are in the X-Z plane with an angle $\theta_i$ to the
$z$-axis, then a bipartite structure with sublattices $A$ and $B$ is
employed to describe the possible periodicity in the ground
state\cite{Schmid,Bruder,Scalettar}. The candidate states include an
easy-plane AF phase ($\theta_A=-\theta_B$) or paired SF with a
non-vanishing order parameter $\langle f_{i\uparrow}f_{i\downarrow}
\rangle\neq 0$, and a canted AF phase ($\cos\theta_A\neq
\cos\theta_B$), which is actually a checkboard solid with a
non-vanishing $\langle f_{i\uparrow}f_{i\downarrow}\rangle$ in one
sublattice while a vanishing $\langle
f_{i\uparrow}f_{i\downarrow}\rangle$ in another sublattice. In
addition, there is a phase separation (PS) regime caused by the
attractive NN interactions. The easy-plane ferromagnetic phase
($\theta_A=\theta_B$) is proved to have higher energy than the
easy-plane AF phase and does not appear in this system.

\section{mean-field results in a square lattice}
We are now ready to explore the superfluidity of fermionic atoms in
a square optical lattice ($z=4$). Hereafter we use the units of
$U=1$. Figure 1 displays the $V-\nu$ phase diagram for hopping
integrals $t_\downarrow=0.1$. We take, e.g., the ratio
$t_\uparrow/t_\downarrow=1.1$. Other choice of
$t_\uparrow/t_\downarrow$ does not alter the conclusion
qualitatively. We compare the mean-field energies of each candidate
phases to determine the ground state. Generally, the canted canted
AF phase with $\cos\theta_A\neq \cos\theta_B\neq 0,1$ is a
supersolid. But in the square lattice we find $\theta_B=0$ or $\pi$,
implying the supersolid order degenerates to an ordinary solid. We
will reexamine this issue through the low energy excitations. The
solid phase takes place in the regime of $0.4\lesssim\nu\lesssim
0.6$. For the SF order, there is a $\pi$-phase difference between
the two sub-lattices (canted AF order). The phase diagram is
symmetrical with respect to the pair-filling $\nu=0.5$ which results
from particle-hole symmetry of the effective Hamiltonian
(\ref{eff1}).

\begin{figure}[tbh]
\includegraphics[width=8cm]{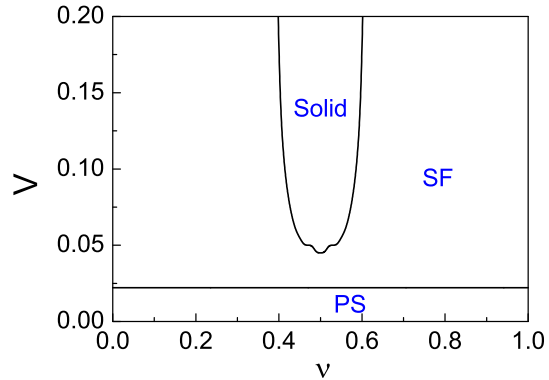}
\caption{The $V-\nu$ phase diagram for a square optical lattice
system with $t_{\downarrow}=0.1$ and
$t_{\uparrow}=1.1t_{\downarrow}$.}
\end{figure}

We discuss the superfluidity in terms of the condensate density
$\rho_{si}=|\langle S_i^-\rangle|^2=|\langle f_{i\uparrow}
f_{i\downarrow}\rangle|^2$. The SF phase has a uniform condensate
density $\rho_s=\nu(1-\nu)$ for a given pair-filling, independent of
the value of $V$. In Fig. 2, we plot the condensate density for the
SF state (dashed curve) as well as the solid state (solid curves)
versus the filling $\nu$ for $V=0.1$. At the MF level, the
condensate density vanishes in one sublattice while does not vanish
in another sublattice. This indicates that this phase is a usual
checkboard solid instead of a supersolid.  More accurate
calculations such as quantum Monte Carlo simulations have
demonstrated that supersolid states indeed do not exist in a square
lattice system\cite{Wessel,Yunoki}.

\begin{figure}[tbh]
\includegraphics[width=8cm]{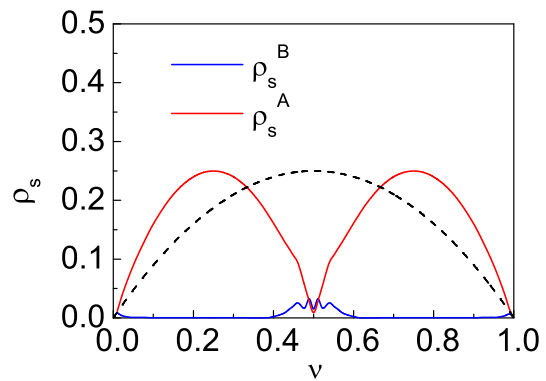}
\caption{(Color online) Condensate density versus the pair-filling
$\nu$ for the superfluid state (dashed curve) and the solid state
(solid curves). For the solid state, the two branches correspond to
sublattices A and B.}
\end{figure}

The superfluidity can also be investigated through the low-energy
excitations, which provides an accurate probe for the nature of the
quantum phase. We study the low-energy excitations by introducing
pseudo-spin operators $a_i^\dagger (a_i)$ for sublattice $A$ and
$b_i^\dagger (b_i)$ for sublattice $B$, respectively. In this case,
$a_i^\dagger=S_i^+=f_{i\downarrow}^\dagger f_{i\uparrow}^\dagger$,
such as for sublattice A, is a composite bosonic operator. After
making a rotation to align the local spins along the $z$-direction,
we obtain the spin-wave type Hamiltonian in the momentum space as,
\begin{widetext}
\begin{eqnarray}
H_{\textrm{sw}}&=&\sum_{\textbf{k}}\{\frac{\gamma_\textbf{k}}{2}
(2\tilde{t}
\cos\theta_A\cos\theta_B+\tilde{U}\sin\theta_A\sin\theta_B)
(a_\textbf{k}^\dagger b_\textbf{-k}^\dagger+a_\textbf{k}
b_\textbf{-k} +a_\textbf{k}^\dagger b_\textbf{k}+a_\textbf{k}
b_\textbf{k}^\dagger)-\tilde{t}\gamma_\textbf{k}(a_\textbf{k}^\dagger
b_\textbf{-k}^\dagger+a_\textbf{k}b_\textbf{-k}-a_\textbf{k}^\dagger
b_\textbf{k}-a_\textbf{k} b_\textbf{k}^\dagger)\nonumber \\
&-&(2\tilde{t}\sin\theta_A\sin\theta_B+\tilde{U}\cos\theta_A\cos\theta_B)
(a_\textbf{k}^\dagger a_\textbf{k}+b_\textbf{k}^\dagger
b_\textbf{k}) -\frac{\tilde{H_z}}{2}(\cos\theta_A
a_\textbf{k}^\dagger a_\textbf{k}+\cos\theta_B b_\textbf{k}^\dagger
b_\textbf{k})\}+H_{linear},\label{SW}
\end{eqnarray}
\end{widetext}
where $\gamma_\textbf{k}=(\cos k_x+\cos k_y)/2$, the renormalized
parameters $\tilde{t}=4t_\uparrow t_\downarrow/U$,
$\tilde{U}=4[V-(t_\uparrow^2 +t_\downarrow^2)/U]$,
$\tilde{H}_z=2(zV-\mu)$ and the summation for momentum $\textbf{k}$
is restricted to a half of the Brillouin zone. $H_{linear}$ includes
the linear term in operators $a_\textbf{k}$ and $b_\textbf{k}$. With
$H_{linear}=0$, $\theta_A$ and $\theta_B$ are determined and the
above MF result is recovered. This Hamiltonian (\ref{SW}) can be
diagonalized in terms of the Bogoliubov transformation to obtain the
low-energy excitation spectrum. For the SF state we have,
\begin{equation}
\omega^2_\textbf{k}=4\tilde{t}(1-\gamma_\textbf{k})
[\tilde{t}(1-\gamma_\textbf{k})
+2\nu(1-\nu)(\tilde{U}+2\tilde{t})\gamma_\textbf{k}].
\end{equation}

\begin{figure}[tbh]
\includegraphics[width=8cm]{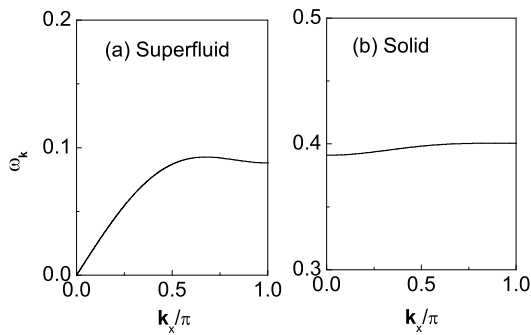}
\caption{The excitation energies of (a) the superfluid state at
$\nu=0.2$ and (b) the solid state at $\nu=0.45$ for the momentum
along the $x$-direction. $t_{\downarrow}=0.1$,
$t_{\uparrow}=1.1t_{\downarrow}$, and $V=0.1$.}
\end{figure}

Figure 3 exhibits the excitation energies versus the momentum along
the $x$-direction for $t_{\downarrow}=0.1$,
$t_{\uparrow}=1.1t_{\downarrow}$, and $V=0.1$. For the superfluid
state in Fig.3(a) with $\nu=0.2$, the energy spectrum is linear at
small momentum which reveals a gapless mode. At larger momenta, a
energy dip plays the role of a roton as in the $^4 He$ superfluid.
The excitations of the solid state in Fig.3(b) with $\nu=0.45$
reveals a gapful mode. No gapless low energy excitation mode is
found in this parameter regime. It justifies that this phase is a
usual checkboard solid rather than a supersolid.

\section{Spontaneous superflow in a triangular lattice}
We now study an intriguing phenomenon of spontaneous superflow of
the fermionic superfluidity for a triangular optical lattice system.
Although there are some debates on the possible disordered ground
state in a triangular AF model because of the geometric frustration
as well as quantum fluctuations, the current consensus is that the
ground state has a long-range N\'{e}el order
\cite{Huse,Nishimori,Jolicoeur,Capriotti}. We consider the classical
spins $\textbf{S}_i$ and employ the $120^0$ XY-N\'{e}el order for
three sublattices A, B, and C. Let $\theta_A$, $\theta_B$ and
$\theta_C$ be the corresponding polar angles which reflect the
spatial density variations\cite{Schmid,Bruder,Scalettar}, the MF
energy of the system is written as
\begin{widetext}
\begin{eqnarray}
E_{{\rm MF}}&=&\frac{3}{2}(V-\frac{t_\uparrow^2+t_\downarrow^2}{U})
(\cos\theta_A\cos\theta_B+\cos\theta_B\cos\theta_C+
\cos\theta_C\cos\theta_A)\label{MF}\\\nonumber
&&-\frac{3t_\uparrow
t_\downarrow}{2U}(\sin\theta_A\sin\theta_B+\sin\theta_B\sin\theta_C
+\sin\theta_C\sin\theta_A)+(V-\frac{\mu}{3})
(\cos\theta_A+\cos\theta_B+\cos\theta_C),
\end{eqnarray}
\end{widetext}
where $\mu$ is the Lagrangian multiplier that controls the total
pair-filling. In formula (\ref{MF}), the $2\pi/3$ azimuthal angle
difference between spins in the three sublattices has been
incorporated.

The MF energy should be minimized with respect to the angles
$\theta_A$, $\theta_B$, and $\theta_C$ at a given polarization
$\langle S_z\rangle=\nu-\frac{1}{2}$. It has a form similar to that
in Ref. [\onlinecite{Murthy}] except the first term may become
negative. In that case, the system is phase separated. Generally,
there is a supersolid phase with $\theta_A=\theta_B\neq \theta_C$.
For $V<(t_\uparrow+t_\downarrow)^2/U$, the ground-state is a uniform
superfluid. We focus on the uniform superfluid phase
($\theta_A=\theta_B=\theta_C=\cos^{-1}(\nu-\frac{1}{2})$), which is
implemented at a moderate NN interaction $V\gtrsim (t_\uparrow^2+
t_\downarrow^2)/{U}$. We explore the implications of this $120^0$
N\'{e}el state and its possible consequence in the paired
superfluid.

\begin{figure}[tbh]
\includegraphics[width=8cm]{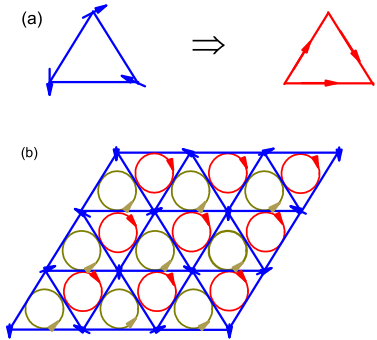}
\caption{(Color online) Spontaneous superflows of the paired
superfluid in the optical triangular lattice. (a) The azimuthal
angles between spins correspond to the phase difference between
superfluids on neighboring sites and a Josephson superflow
spontaneously occurs along the edge. (b) The arrowed circles
represent vortex-antivortex pairs which form the superflow
convection.}
\end{figure}

According to the spin mapping $\langle f_{i\uparrow}f_{i\downarrow}
\rangle=\langle S_i^-\rangle=\frac{1}{2}\sin\theta_i e^{-i\phi_i}$,
the paired superfluid has an order parameter phase as that of the
azimuthal angle $\phi_i$ of the spin. Therefore, the $2\pi/3$
azimuthal angle difference of the spins implies a $\Delta
\phi_{ij}=2\pi/3$ phase difference between neighboring sites of the
superfluid. The N\'{e}el order of the antiferromagnetic model thus
corresponds to a periodical phase modulation in the superfluid. In
the theory of the Josephson tunneling, two weakly connected
superfluids or superconductors will induce a current as a result of
their phase difference as $J\propto \sin\Delta\phi$. Consequently,
the ground state superfluid spontaneously flows along the link of
the neighboring sites, as is shown in Fig.4(a). In the triangular
lattice system, the superfluid flows form a closed ring-like vortex.
Figure 4(b) schematically displays a regular convection pattern of
superflows which contains a sequence of the vortex-antivortex
pairs.

Similar cellular superflows and periodic textures were suggested in
the $^3He-A$ superfluid when a perpendicular magnetic field is
applied to a sample slap\cite{Hu}. The coupling between the
superfluid velocity and the orbital axis favors spontaneous
superflows. Early theoretical discussions of possible superflow in
solid $^4He$ involved quantum tunneling through ground-state
vacancies, as well as Bose-Einstein condensation and quantum
exchanges within the lattice\cite{Andreev2,Chester,Leggett}. In
2004, Kim and Chan reported the observation of the unusual superflow
without resistance from frictional forces in crystalline
helium\cite{Kim1,Kim2}. This remarkable finding has now been
confirmed\cite{Kim3,Rittner,Sasaki}. The latest experiments indicate
that, rather than being an intrinsic property of a perfect quantum
solid, superflows owe their existence to macroscopic defects or
extended disorder in the structure of solid helium.

In a mismatched Josephson junction of ultracold fermionic atomic
gases, M. L. Kuli\'{c}\cite{Kulic} proposed an oscillating
superfluid amplitude inside the weak link and as a result the
so-called $\pi$-junction. If the junction is a part of the closed
ring then spontaneous and dissipationless superfluid current can
flow through the ring.

\section{Summary and discussions}
We have studied the superfluidity of strongly repulsive fermionic
atoms from a correlated pairing mechanism. The superfluid is a
metastable state with the optical lattice sites either doubly
occupied or empty. The composite objects transport in the optical
lattice through the second quantum processes via virtual
pair-breaking states. It exhibits superfluidity below a critical
temperature. Phase diagrams and low-energy excitations in the square
optical lattice system are investigated. Due to the AF frustration,
the correlated pairs may result in an appealing spontaneous
superflow phenomenon in the triangular optical lattice system.

Some authors explored the possibility of formation of the non-s-wave
BEC through Feshbach resonance in a nonzero angular momentum channel
on a lattice with double occupation\cite{Girvin,Liu,Kuklov3}.
Varying the detuning of non-s-wave resonance can lead to various
quantum phase transitions between the phases: S-wave BEC, non-S-wave
BEC, conventional Mott insulator and orbital Mott insulator (with
broken lattice symmetries). This becomes possible when the atoms are
confined in the $p$-orbital Bloch band of an optical lattice rather
than the usual $s$-orbital band. The new condensate simultaneously
forms an order of transversely staggered orbital currents,
reminiscent of orbital antiferromagnetism or d-density wave in
correlated electronic systems but different in fundamental ways.

The NN interaction depends on the overlap of the Wannier functions
between the NN sites. A moderate value of $V\sim (t^2_\downarrow+
t^2_\uparrow)/U$ is sufficient to create the superfluidity in the
system. An alternative way of generating NN interaction by the
long-range dipolar interaction is also
possible\cite{Goral,Griesmaier}. In order to detect the
superfluidity of the correlated pairs, photoassociation spectroscopy
may be used\cite{Wynar}. Interference of matter waves released from
the lattice has been used to probe the superfluidity of single atom
condensation\cite{Greiner}. By tuning the interaction from repulsive
into attractive, the fermionic atom pairs are converted into
molecules. The sharp peaks will appear in the interference pattern
of the released bosonic molecules due to the presence of a SF
fraction.

This work is supported by the National Natural Science Foundation of
China under grant No. 10874018 and by the 973 Program Project under
grant No. 2009CB929101.

\end{document}